\documentclass[aps,prl,twocolumn,nobibnotes]{revtex4}
\usepackage{graphics,graphicx,amsfonts,amsmath,amsbsy,amssymb,color}
\usepackage{bm}
\usepackage{subfigure}
\usepackage{multirow}
\usepackage{vector}  

\def \beq {\begin{eqnarray}}
\def \eeq {\end{eqnarray}}

\def \mEh {{\textrm{mE}_{\textrm{h}}}}
\def \Eh {{\textrm{E}_{\textrm{h}}}}

\def \nadd {{n_{\textrm{add}}}}

\newcommand{\rff}[1]{{Eq.~\eqref{#1}}}

\def \wftrial {{ |\psi_{\textrm{T}} \rangle }}

\begin{document}
\title{Excited States, Dynamic Correlation Functions and Spectral Properties from Full Configuration Interaction Quantum Monte Carlo}
\author{George~H.~Booth}
\author{Garnet~Kin-Lic~Chan}
\affiliation{Department of Chemistry, Frick Laboratory, Princeton University, NJ 08544, USA}

\begin{abstract}
In this communication, we propose a method for obtaining isolated excited states within the Full Configuration Interaction Quantum Monte Carlo framework.
This method allows for stable sampling with respect to collapse to lower energy states and requires no uncontrolled approximations.
In contrast with most previous methods to extract
excited state information from Quantum Monte Carlo methods, this results from a modification to the underlying 
propagator, and does not require explicit orthogonalization, analytic continuation, transient estimators or 
restriction of the Hilbert space via a trial wavefunction. Furthermore, we show that the propagator can directly yield 
frequency-domain correlation functions and spectral functions such as the density of states which are difficult to obtain within a traditional Quantum Monte Carlo
framework.
We demonstrate this approach with pilot applications to the neon atom and beryllium dimer.
\end{abstract}
\date{\today}
\maketitle

Almost all of the many variants of projector Quantum Monte Carlo (QMC) 
rely on the properties of the operator $e^{-\beta H}$, where due to its
similarity to the time-evolution operator, the variable $\beta$ is denoted `imaginary time'. 
Generally, this imaginary time is discretized, and the operator iteratively applied as a short-time propagator, 
in order to simulate its action in the large $\beta$ limit\cite{Foulkes2001}.
Expressing an initial wavefunction in the eigenbasis of the Hamiltonian of interest, the application of this $e^{-\beta H}$ propagator results in a 
projection onto the $i^{\textrm{th}}$ eigenstate proportional to $e^{-\beta E_i}$, where $E_i$ is the energy of this eigenstate. It is clear to see
that in this large $\beta$ limit, the projection onto the lowest energy eigenvector dominates the wavefunction, whereby ground state 
properties can be extracted, assuming some overlap with the initial wavefunction. While this formalism is clearly powerful, by construction it 
exponentially quickly projects out excited states of the system which may be of interest, 
and are of critical importance in the simulation of finite-temperature properties, reaction dynamics, photochemistry and many other areas. 

To date, isolating excited states of systems via projector QMC methods has only been 
practical with a restriction on the projection to sample a space which is approximately orthogonal to those of the lower energy states, via nodal constraint\cite{Lester1986}, or
orthogonalization against them in a subspace projection method\cite{Ceperley1988,Nagase2010}. More indirectly, statistical methods have been used on short periods of 
imaginary-time in order to isolate individual decay rates in the spectrum by analytic continuation to a real-time dynamic\cite{Baym1961,Gubernatis91}. However, these approaches are not entirely 
satisfactory; accurate nodal surfaces for excited states can be difficult to obtain, resulting in a larger fixed node error and potentially transient
estimators, while the subspace projection method has limited applicability\cite{Ceperley1997}. In addition, the Bayesian techniques which rely on maximizing entropy to
approximate a notoriously unstable inverse Laplace transform, have difficulty achieving quantitative accuracy within noisy data sets \cite{Berne1983,Blume1997,Gunnarsson10}. 
Despite this, there are examples of accurate excited states within the nodal constraint\cite{Umrigar2009}, while maximum 
entropy techniques are particularly prevalent in solid state calculations to obtain the density of states, often in the case of continuous-time QMC 
as applied to quantum impurity models and dynamical mean-field theory\cite{Liu2012,Millis2011}.

Here, we take a different approach to the calculation of excited states, within the context of Full Configuration Interaction Quantum 
Monte Carlo (FCIQMC)\cite{BTA2009,CBA2010,BoothC2}. This recently introduced method applies
the imaginary-time evolution propagator to a stochastic `walker' representation of the wavefunction expressed in the full space of Slater 
determinants constructed from a single-particle basis of size $M$. Although this
reintroduces a basis set error compared to those methods operating in real space, it confers various advantages which mitigate this.
The discrete basis allows for an efficient walker annihilation algorithm, which can exactly overcome the fermion 
sign problem in the sampling, provided enough walkers are used\cite{Spencer2012}. The `initiator' approximation was 
formulated to maintain a high annihilation rate, and control the growth of noise in a systematically improvable fashion\cite{CBA2010,BoothC2}. This has allowed 
far larger systems to be treated at an accuracy
comparable to that of Full Configuration Interaction (FCI or exact diagonalization), within small and systematically improvable error bars. In addition, a semi-stochastic adaptation 
of the algorithm\cite{UmrigarArXiv}, as well the introduction of a partial nodal constraint\cite{Clark2012} and ideas from quantum chemistry\cite{BoothF12} hold promise of increased accuracy and efficiency
of the method.

In order to project out a targeted excited state rather than the ground state, we propose the use of a projection operator of the form
\begin{equation}
P(H)=e^{-\beta^2(H-S)^2} . \label{eqn:Propagator} 
\end{equation}
For sufficiently large $\beta$, this Gaussian propagator will result in the dominant eigenstate being
the one closest in energy to the chosen value of the diagonal offset $S$, termed the shift.
In the eigenbasis of $H$, $\{ |\Psi_i\rangle, E_i \}$ with $\Psi_0$ representing the ground state, and starting from an initial wavefunction $\wftrial$ with $S=E_k$, 
it can be seen that the long time propagation results in
\begin{equation}
|\Psi_k\rangle \propto \lim_{\beta\rightarrow\infty} \sum_i |\Psi_i\rangle e^{-\beta^2(E_i-E_k)^2} \langle \Psi_i \wftrial \propto \sum_i \delta_{E_i,E_k} |\Psi_i \rangle . 
\end{equation}
We note here that a projector of this form was proposed back in 1983 within continuum QMC approaches\cite{Hirsch1983,Berne1985}, although due to sign problems, and significantly 
larger timestep errors resulting from the fact that $H^2$ is more singular than $H$, only one-electron systems were reported, and no modern implementation exists
in the literature.
This issue of the timestep highlights another advantage of working in a finite basis, in that the spectrum is bounded both from below and above. This allows
for linearization of the short-time propagator, 
\begin{eqnarray}
|\Psi\rangle &\propto& \lim_{P\rightarrow\infty} \left[A e^{-\tau (H-S)^2}\right]^P \wftrial \\
|\Psi\rangle &\propto& \lim_{P\rightarrow\infty} \left[A(1-(\tau (H-S))^2)\right]^P \wftrial  ,
\end{eqnarray}
without becoming unbound and dominated by very high energy states oscillating in time, and without incurring timestep errors in the final wavefunction so long as the
timestep is less than an upper bound given by $\tau \leq \frac{2}{E_{\mathrm{max}}-E_{\mathrm{min}}}$. 
Repeated application of the short-time propagator therefore results in a 
`power-method' for states on the interior of the spectrum, rather than at the extrema, with similarities to filter diagonalization\cite{Grosso1993,Neuhauser95}.

Propagation with \rff{eqn:Propagator} leads to a theoretical decay of state $j$ from $i$ as $e^{-\beta^2((E_i-S)^2-(E_j-S)^2)}$. In contrast with the ground state propagator, this
rate of decay depends on $S$, with it being advantageous to choose $S$ to be as close to the energy of the state of interest as possible.
However, even if $S$ is chosen exactly, the projection of the non-dominant states is slower compared to the ground state propagation, and we will return to this issue later. 
In addition, unless $S$ is chosen exactly, the long-time propagation 
of the dynamic will result in a continued projection onto a decaying function of all states, including the dominant one. For this reason, the factor of $A$ is introduced into the short-time
propagator, such that
at convergence, its value can be varied in order to maintain a constant L$^1$ normalization of the dominant wavefunction and a stable number of walkers. This is analogous to the variation of $S$
in the ground state projection\cite{BTA2009}.

The differential formulation of the exact dynamic governed by this propagator for a given component of the wavefunction, $C_i$, can be written as
\begin{equation}
\frac{d C_i}{d \tau^2} = (A-1)C_i - \epsilon A \sum_{j,k} (H_{ij}-\delta_{ij}S)(H_{jk}-\delta_{jk}S) C_k    ,   \label{eqn:Dynamic}
\end{equation}
where $\epsilon \rightarrow \tau^2$ as $A \rightarrow 1$, and the application of $H^2$ has been decomposed by a resolution of the identity over the connecting space of determinants $j$.
This formulation is now amenable to stochastic integration with a discrete walker representation of the determinantal wavefunction coefficients $C$. As with the ground state projection, there is no
unique stochastic algorithm for this dynamic, but the one which we found to be most efficient involves a double spawning cycle, which requires little additional overhead compared to the ground state
algorithm, and no additional memory requirements. Each iteration, the determinants represented by $k$ are 
run through, and a spawning step attempted to determinant $j$, in the same fashion as the ground state propagation, but in negative time. This results in a spawning probability
to a connected determinant $j$ with a stochastically realised {\em signed} amplitude of $\frac{\tau H_{jk}}{P(k|j)}$, where $P(k|j)$ represents the normalised probability to randomly select symmetry-connected
determinant $j$ from $i$. Successfully spawned walkers are subsequently propagated again in the same iteration with a now forwards-time signed amplitude 
of $-\frac{\tau H_{ji}}{P(j|i)}$. Care must be taken that for determinant $k$, the diagonal `death' processes from the first application of $H$ are now interpreted as spawning events, 
which are also subsequently propagated
via $H_{ij}$. Each iteration, the factor of $A$ is applied initially as a separate enhancement of the local population of each determinant, with the absolute population on each
determinant growing with probability $A C_k$.

\begin{figure}[t]
\begin{center}
\includegraphics[scale=0.475]{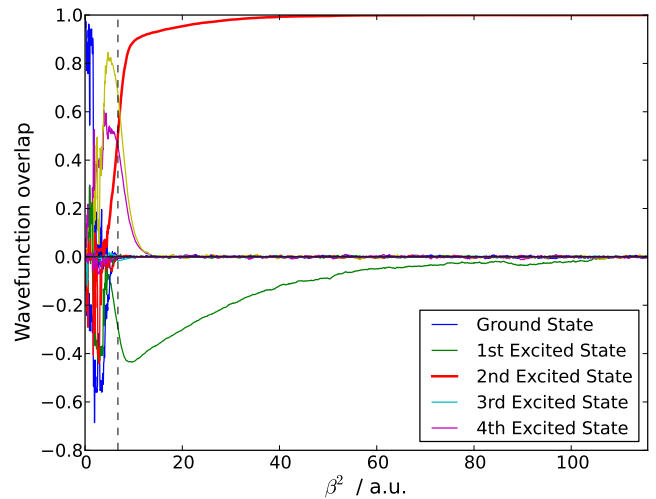}
\end{center}
\caption{Convergence of the propagation to the second excited state of He$_2$ at 2.5~\AA\ separation in a cc-pVDZ basis. 
$S$ was fixed at -3.65$\Eh$, and $A$ at 1.004 until 10,000 walkers were present, denoted by the dotted line, 
where $A$ was varied in order to keep this number constant. After variation, the average value of $A-1$ was 1.4(6)$\times10^{-6}$.
}
\label{He2conv}
\end{figure}

\begin{figure}[t]
\begin{center}
\includegraphics[scale=0.475]{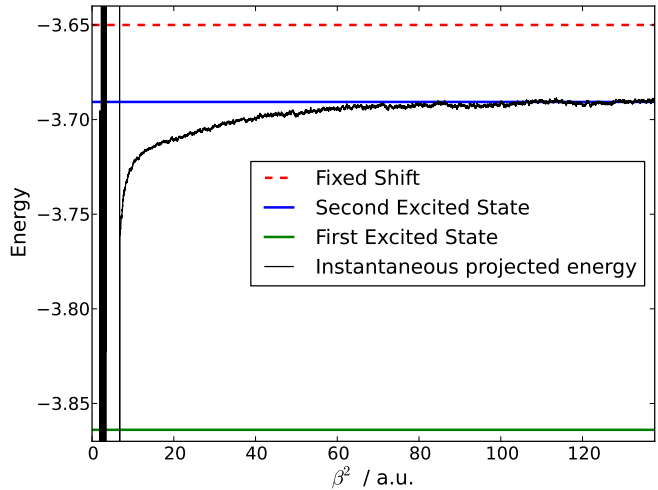}
\end{center}
\caption{
Convergence of the projected energy estimate to the exact eigenvalue. The reference determinant for the projection was dynamically adjusted to project onto the largest
weighted determinant in the sample.
}
\label{He2Energy}
\end{figure}

In Fig.~\ref{He2conv}, we present an illustrative example of the algorithm for the helium dimer in a cc-pVDZ basis, small enough such that the full spectrum of eigenstates can be 
calculated and the convergence of the method analysed. The value of $S$ was fixed at $\sim40\mEh$ higher than the second excited state, but such that it remained the dominant state in the dynamic,
and was subsequently found to be correctly projected out over time. This is despite working in a canonical representation, and starting with a single walker on the Hartree--Fock
determinant which had an initial overlap with the ground state of close to one. In order to grow the walkers, $A$ was initially fixed at a value of 1.004, and was varied when a target number of
walkers was reached, in common with the procedure for the ground state propagation. The convergence of the projected energy, as defined in Ref.~\onlinecite{BTA2009}, is 
shown in Fig.~\ref{He2Energy} for the same simulation, and reflects the decay from the sampled wavefunction of the first excited state.

In order to reliably extend to larger molecular systems, it is worth considering how to transfer the salient elements of the initiator approximation into this new propagation. 
The basis of this approximation is to attempt to propagate walkers corresponding to wavefunction signal exactly, while walkers judged to be potentially noise are propagated with a truncated
Hamiltonian which acts only over the instantaneously occupied subspace\cite{CBA2010,BoothC2}. This is systematically improvable as the instantaneously occupied subspace grows, or the criterion
for walkers corresponding to signal becomes more inclusive.
The separation between walkers corresponding to signal and potential noise is not unique.
However, it seems sensible to retain the tested feature from ground state propagation that newly spawned walkers on previously unoccupied determinants ($i$) at the end of an iteration, must 
have come from an initial determinant ($k$) which is deemed to have a well-established sign, and therefore a population of walkers above $\nadd$. Consequently, all walkers from the application
of the first Hamiltonian operator are kept, while the information as to whether $C_i$ is above the initiator criterion is passed through to the annihilation stage of the final set of spawned walkers.
No walkers are therefore aborted over the resolution of the identity between $H^2$ (determinants $j$ in \rff{eqn:Dynamic}).

To test this on a larger system, we study an excited state deep in the spectrum of the 10-electron neon atom in a cc-pVDZ basis, with an energy of approximately $2.5\Eh$ above the ground state.
Setting $S$ to equal the CISDTQ energy for the corresponding state, and while remaining in a canonical Hartree--Fock basis and starting from a random distribution of walkers throughout the whole space, 
we achieved a converged energy of -126.2118(4)$\Eh$, compared to the FCI value of -126.21177$\Eh$. This value is $4.76\mEh$ lower than the initial guess provided by CISDTQ.
It would be highly advantageous to develop a robust algorithm for varying $S$ dynamically during the run, as is done for the ground 
state algorithm. This could be used to maximise the rate of growth of walkers or alternatively minimize $A$, 
both of which should adjust $S$ to more closely match the eigenvalue of the state, remove reliance on the initial guess and increase the convergence rate. 
However, since this requires finding a minimum in a quadratically varying and noisy dataset, no robust algorithm has been identified so far.

Dynamic correlation and response functions due to some perturbation, either in the frequency or time domain, are of critical importance in electronic structure theory\cite{Balseiro1987}, and 
are directly measured in experiments to probe the electronic properties of materials through techniques such as neutron scattering or photoelectron spectroscopy\cite{Hutchings1972}. Many methods, 
including in general QMC approaches, have significant difficulty in calculating these quantities\cite{Berne1991}, often having to rely on unstable analytic continuation from 
imaginary time correlation functions\cite{Baym1961,Berne1983,Gubernatis91,Gunnarsson10,Blume1997}, while  
other methods can bias towards high or low energy regimes\cite{Millis2011}. 
Although other correlation functions are possible, here we look at the example of an advanced Green's function, a central concept in electronic structure 
where the `perturbation' at time $t=0$ is the creation of a hole in orbital $j$. For negative time periods, $t$, these can be written in the time and frequency domain respectively as
\begin{eqnarray}
G^-(i,j,t) &=& i\langle \Psi_0|a^{\dagger}_i e^{-i(H-E_0-i\delta)t} a_j| \Psi_0 \rangle  \\
G^-(i,j,\omega) &=& \langle \Psi_0|a^{\dagger}_i \frac{1}{\omega-(H-E_0)+i\delta} a_j| \Psi_0 \rangle .  \label{eqn:FreqGF}
\end{eqnarray}
Unlike the inverse Laplace transform required for the analytic continuation of imaginary time correlation functions, the transform between these two domains is a well-conditioned and numerically stable 
fourier transform in the presence of noisy data.
Spectral density functions, such as the density of states for extended systems, are then defined in the Lehmann representation\cite{FetterandW} as
\begin{eqnarray}
A^-(i,j,\omega) &=& -\frac{1}{\pi} \Im[G^-(i,j,\omega)]  \\
                &=& \frac{1}{\pi}\sum_{n}\frac{\langle \Psi_0^{N}|a^{\dagger}_i|\Psi_n^{N-1}\rangle \delta \langle \Psi_n^{N-1} | a_j | \Psi_0^{N} \rangle}{(\omega-E_n^{N-1}+E_0^N)^2+\delta^2}  , \label{eqn:ShortTimeSpectral}
\end{eqnarray}
which in the small $\delta$ limit tends to
\begin{align}
A^-(i,j,\omega) =& \sum_{n}\langle \Psi_0^{N}|a^{\dagger}_i|\Psi_n^{N-1}\rangle \langle \Psi_n^{N-1} | a_j | \Psi_0^{N} \rangle \times \notag \\
                 &\delta(\omega-(E_n^{N-1}-E_0^N)) ,  \label{eqn:Spectral}
\end{align}
where $\delta$ in the above equation represents the dirac-delta function.

\begin{figure}[t]
\begin{center}
\includegraphics[scale=0.475]{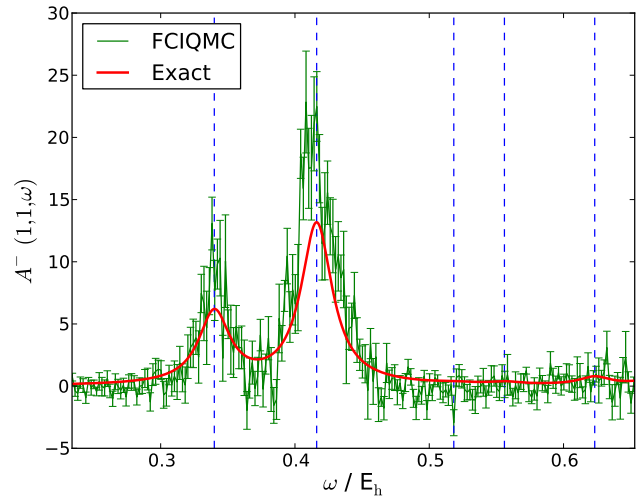}
\end{center}
\caption{
High energy window of the spectral function $A^-(1,1,\omega)$ for exact propagation with $\delta=0.0141\Eh$, and stochastic evaluation 
via FCIQMC for an equivalent time $\beta=50$a.u. for frozen-core Be$_2$ in a cc-pVDZ basis at 2.5\AA\ . 
Vertical lines indicate the difference between the ground state energy and the eigenvalues of the N-1 system symmetry connected in $G^-(1,1,\omega)$, although some are coupled 
too weakly to contribute significantly to the spectral function. Approximately 10 independent calculations at each value of $\omega$ were averaged to obtain the errorbars.
}
\label{Beryllium}
\end{figure}

Assuming $A=1$, application of the propagator in \rff{eqn:Propagator} for a time $\beta^2=\frac{1}{2 \delta^2}$ will result in the wavefunction
\begin{equation}
C(\beta^2)=e^{-\frac{1}{2 \delta^2}(H-S)^2}\wftrial  ,   \label{eqn:IntDynamic}
\end{equation}
which when applied to an initial wavefunction $\wftrial=a_j|\Psi_0 \rangle$ obtained from the ground-state dynamic, and then projected 
onto $\frac{\beta}{\sqrt{\pi}}\langle \Psi_0 |a^{\dagger}_i$ will result in the distribution
\begin{align}
f(i,j,\omega) =& \frac{1}{\sqrt{2 \pi} \delta} \sum_{n}\langle \Psi_0^{N}|a^{\dagger}_i|\Psi_n^{N-1}\rangle \langle \Psi_n^{N-1} | a_j | \Psi_0^{N} \rangle \times \notag \\
               & e^{-\frac{1}{2 \delta^2}(\omega-E_n^{N-1}+E_0^N)^2}    \label{eqn:GF}
\end{align}
for $S=\omega+E_0^N$. This will tend to the spectral function given in \rff{eqn:Spectral} in the large $\beta$ limit.
The real parts of the Green's function can then by obtained if needed 
from the Kramers-Kronig relation\cite{Balseiro1987}.
We note that a related Green's function can be obtained directly by integrating \rff{eqn:IntDynamic} over $\beta$, with the addition of the small
imaginary component $\delta$ to the dynamic. Unfortunately however, this integral is only convergent for $(\omega-E_n^{N-1}+E_0^N) > \delta$, and so 
the FCIQMC calculation will blow up at the poles. In systems with a continuous spectra, this would not be appropriate, and so we do not pursue this approach here.
Results from a pilot investigation of the beryllium dimer in a cc-pVDZ basis, where the exact Green's function can be obtained from complete diagonalization, are shown in Fig.~\ref{Beryllium}.

In order to reduce the statistical error, it may be necessary to
average over a small number of independent calculations at each frequency, and this can be combined with an elimination of the bias derived from choosing a correlated sample 
of $\langle \Psi_0|a^{\dagger}_i$ and $a_j| \Psi_0 \rangle$ \cite{Ceperley1988}, by taking the $\Psi_0$ samples on each side of \rff{eqn:GF} from 
different snapshots in imaginary time.
In addition, by storing multiple wavefunctions of the type $\langle \Psi_0|a^{\dagger}_i$ at the same time, all $M^2$ single-particle Green's 
functions can be calculated at a cost of $\mathcal{O}[M]$ FCIQMC calculations per frequency point, without the expectation of any variation in 
accuracy between high and low energy regimes. 

However, despite modest successes, it is clear that obtaining converged results through the use of this operator is substantially more 
difficult than with the ground state projection.
This is mainly due to an additional factor of $(\tau \Delta E)^{-1}$ in the number of iterations required to project out undesired states with energy gap $\Delta E$ for
comparable accuracy to the ground state propagation. 
The result is that while in the ground state propagation excited states were filtered relatively quickly with 
only isolated convergence issues in the case of near degeneracy\cite{BoothC2}, the number of iterations required for excited state propagation 
are substantially increased, as well as the dynamic being less well-conditioned with respect to walker fluctuations. 
This is also exacerbated by a generally more multiconfigurational wavefunction which increases random
error in the projected energy estimator\cite{BTA2009}.
A more judicious choice of orbital basis and initial conditions optimized for the state of interest, as well as a multireference
projected energy formulation\cite{UmrigarArXiv} would ameliorate many of these issues. In addition, there is the possibility of
preconditioning techniques familiar from iterative diagonalization methods\cite{Davidson1975} being transferred into the stochastic dynamic, as well other
operators, such as $e^{-\beta |H|}$, which should behave more efficiently and allow for extension to larger systems. Research in these
directions is currently under way.
It is clear that alternative propagators within the FCIQMC dynamic holds promise for obtaining accurate excited states.

\section*{Acknowledgements}


\end{document}